\documentclass{ab}

\usepackage{graphicx}
\usepackage{latexsym}

\def\degr{\hbox{$^\circ$}}

\def\arcsec{\hbox{$^{\prime\prime}$}} 

\begin{document}

\title{Detection of an Intergalactic Meteor Particle \protect \\with the 6-m Telescope }
\author{V.L.~Afanasiev$^{1}$ \and V.V.~Kalenichenko$^{2}$ \and I.D. Karachentsev $^{1}$}
\institute{Special Astrophysical Observatory Russian Academy of Sciences,
N. Arkhyz, KChR, 369167, Russia \and Astronomical Observatory, Kyiv Taras Shevchenko University, 3 Observatorna ul.,  Kyiv, 04053 Ukraine}

\offprints{V.L.  Afanasiev, \email{vafan@sao.ru}}
 
\titlerunning{Detection of an Intergalactic Meteor Particle}

\authorrunning{Afanasiev et al.}
 
\date{Received: July 10, 2007/Revised: August 17, 2007}

\abstract{
On July 28, 2006 the 6-m telescope of the Special Astrophysical Observatory of
the Russian Academy of Sciences recorded the spectrum of a faint meteor. We
confidently identify the lines of FeI and MgI, OI, NI and molecular-nitrogen
(N$_{2}$) bands. The entry velocity of the meteor  body into the Earth's
atmosphere estimated from radial velocity is equal to 300~km/s. The body was
several tens of a millimeter in size, like chondrules in carbon chondrites. The
radiant of the meteor trajectory coincides with the sky position of the apex of
the motion of the Solar system toward the centroid of the Local Group of
galaxies. Observations of faint sporadic meteors with FAVOR TV CCD camera
confirmed the radiant at a higher than  96{\%} confidence level. We conclude
that this meteor particle is likely to be of extragalactic origin. The
following important questions remain open: (1)~How metal-rich dust particles
came to be in the extragalactic space? (2)~Why are the sizes of extragalactic
particles  larger by two orders of magnitude (and their masses greater by
six orders of magnitude) than common interstellar dust grains in our Galaxy?
(3)~If extragalactic dust surrounds galaxies in the form of dust (or
gas-and-dust) aureoles, can such formations now be observed using other
observational techniques (IR observations aboard Spitzer satellite, etc.)?
(4)~If inhomogeneous extragalactic dust medium with the parameters mentioned above
actually exists, does it show up in the form of irregularities on the cosmic
microwave background (WMAP etc.)?
}
\maketitle

\section{INTRODUCTION}
Systematic searches for and investigation of meteors produced by the entry of
cosmic bodies into the Earth's atmosphere are conducted with small wide-angle
cameras [\cite{1:Afanasiev_n}]. Slitless spectra yield spectroscopic data for
meteors brighter than 4--5 magnitude and provide information on the chemical
composition of the disintegrating particles [\cite{2:Afanasiev_n}]. An enormous
number of sporadic meteors fainter than 8 magnitude are known to be observed
besides sufficiently bright meteors associated with meteor streams. Observers
on large telescopes often see very faint meteors in the fields of view of their
instruments. In some cases, slit spectra of these meteors can be obtained
[\cite{3:Afanasiev_n}]. In this paper we report the results of our analysis of
the spectrum and radial velocity of a faint meteor that we accidentally
detected in the field of view of the 6-m telescope of the Special Astrophysical
Observatory of the Russian Academy of Sciences.

\section{OBSERVATIONS}

\begin{figure*}[htbp]

\centerline{\includegraphics[width=14cm]{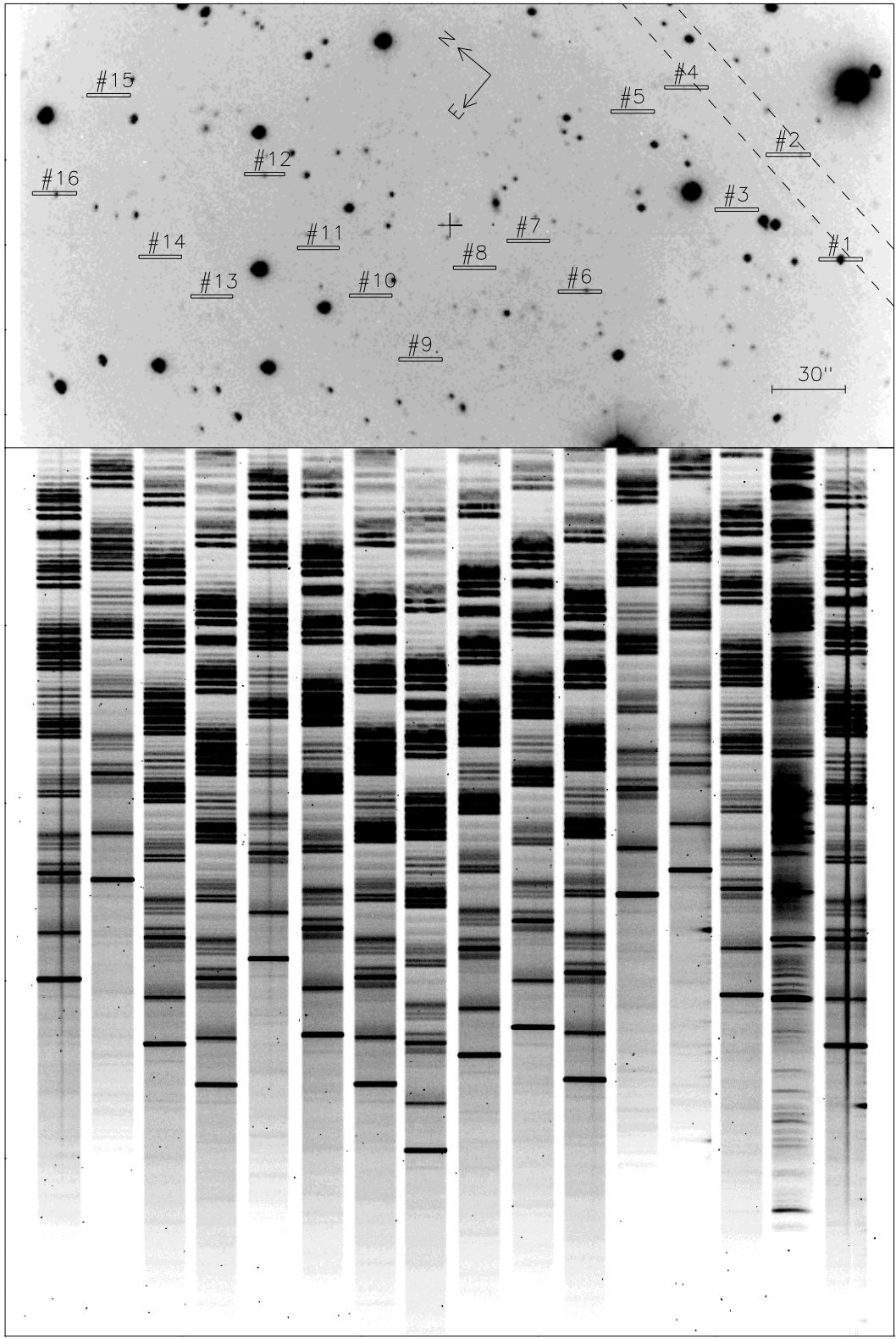}}
 \caption{Spectra in the field 2206+29 taken with SCORPIO
spectrograph operating in multislit mode and attached to the 6-m telescope of
the Special Astrophysical Observatory of the Russian Academy of Sciences. The
upper panel shows the region studied with the slit positions and numbers
marked. The spectrum of the meteor was recorded in slits {\#}1,{\#}2, and
{\#}4. The dashed line shows the hypothetical track of the meteor (its
direction and width).} \label{fig1:Afanasiev_n}
\end{figure*}

\begin{figure*}[htbp]
\includegraphics[width=14cm]{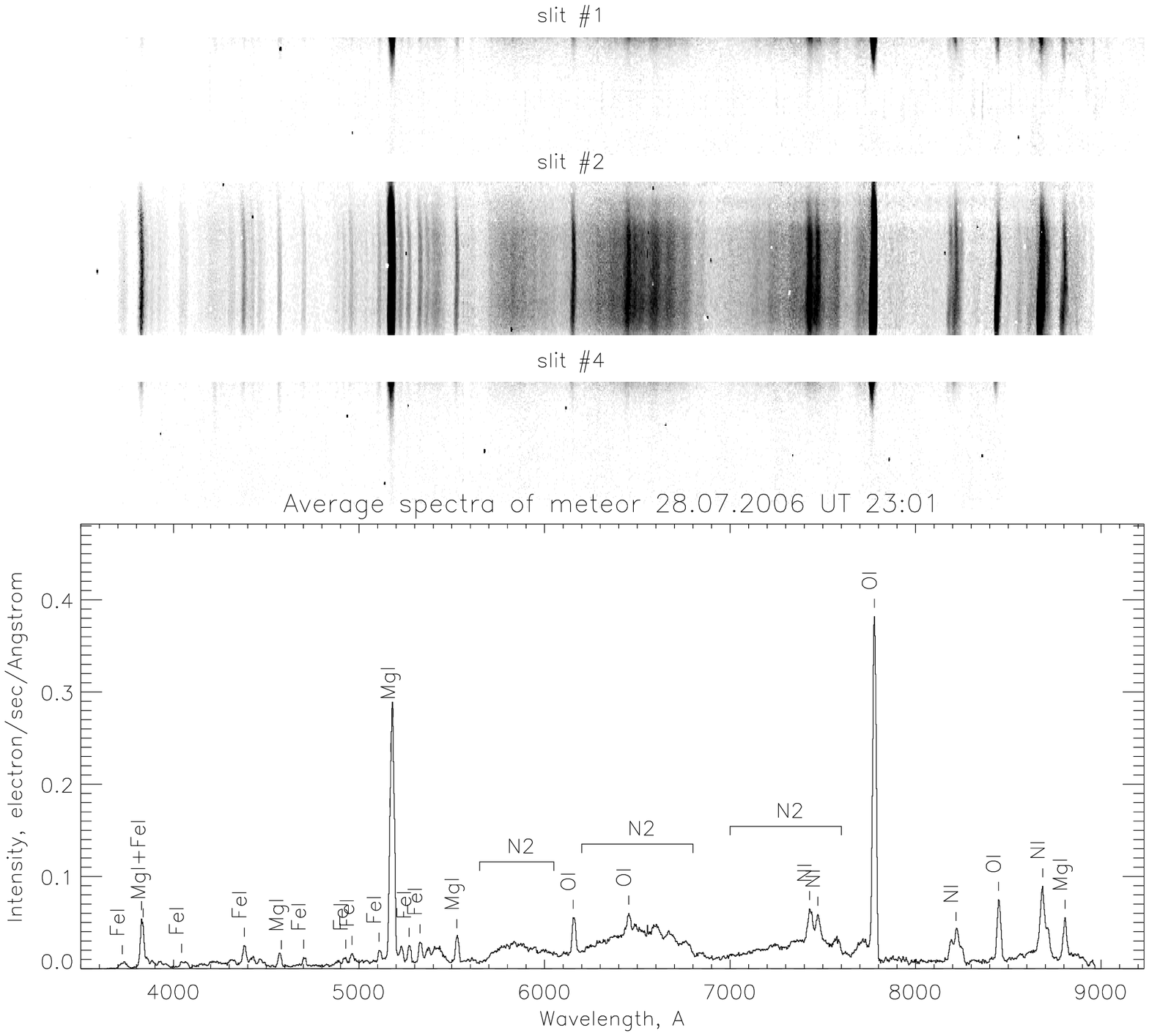}
 \caption{The spectra of the meteor taken in slits
{\#}1,{\#}2, and {\#}4 after subtraction of the sky background and wavelength
calibration. The upper panel shows the images of the spectra in various slits
and the lower panel shows the average spectrum of the meteor for slit {\#}2
with the lines identified.} \label{fig2:Afanasiev_n}
\end{figure*}

We recorded the spectrum of the meteor on July 28, 2006 during our observations
of the spectra of faint galaxies in the  2206+29 field using SCORPIO universal
focal reducer attached to the 6-m telescope of the Special Astrophysical
Observatory of the Russian Academy of Sciences and operating in multislit
spectroscopy mode [\cite{4:Afanasiev_n}]. The spectrograph recorded the
3500--9500\,\AA\AA\ wavelength interval with a spectral resolution of
15--18\,\AA\AA\ via a 2048$\times$2048 EEV42-40 CCD. We use standard
observational technique with the spectrum referred to spectrophotometric
standard  GD248 at a similar zenith distance. We recorded the spectrum of the
faint meteor simultaneously in \mbox{three} slits, thereby allowing the
trajectory of the meteor on the sky to be confidently reconstructed. Figure~1
shows the sky area studied, slit locations, and the corresponding original
images of the spectra. The dashed line shows the hypothetical meteor
trajectory. The upper panel of Fig.\,2 shows the images of the meteor spectra
in three slits after the reduction of scale distortions, wavelength
calibration, and sky-background subtraction. The lower panel shows the average
spectrum of the meteor uncorrected for spectral sensitivity. The FeI and MgI
lines typical of chondrites  [\cite{5:Afanasiev_n}] can be confidently identified
in the blue part of the spectrum, as also evidenced by the relative intensities
of the FeI multiplets and MgI lines  [\cite{6:Afanasiev_n}]. The red part of the
spectrum features conspicuous  OI and NI lines and molecular nitrogen (N$_{2}$)
bands, which are typical of strongly heated air with the temperature of
15000--20000\,K~[\cite{7:Afanasiev_n}]. One can also see the infrared resonance
MgI line, which belongs to the spectrum of the meteor particle proper.

The distortion of lines that shows up conspicuously on the spectra is most
remarkable. It is indicative of the variation of radial velocity along the
slit. The first estimates based on the original spectra as seen in the
telescope yielded radial-velocity variations exceeding 250~km/s, which is
absolutely unusual for meteors. To refine the radial-velocity variations, we
determined the Doppler shifts by cross correlating the spectrum of the meteor
along the slit with the average spectrum calibrated to the laboratory
wavelength scale. The left-hand panel in Fig.~3 shows the isophotes of the
observed radial-velocity distribution (LOSVD) and the right-hand panel, the
Doppler velocities along the slit measured from the position of the maximum. It
is absolutely evident that this pattern corresponds to the spectrum of the
products of evaporation of a meteor particle in the atmosphere --- we see the
deceleration portion corresponding to minimum velocity at the center of the
slit and the portions of maximum velocity corresponding to meteor vapor at the
ends of the spectrum. It follows from this that the velocity of intrusion of
the observed meteor into the Earth's atmosphere was about 300~km/s.

\begin{figure*}[htbp]
\centerline{\includegraphics[width=14cm]{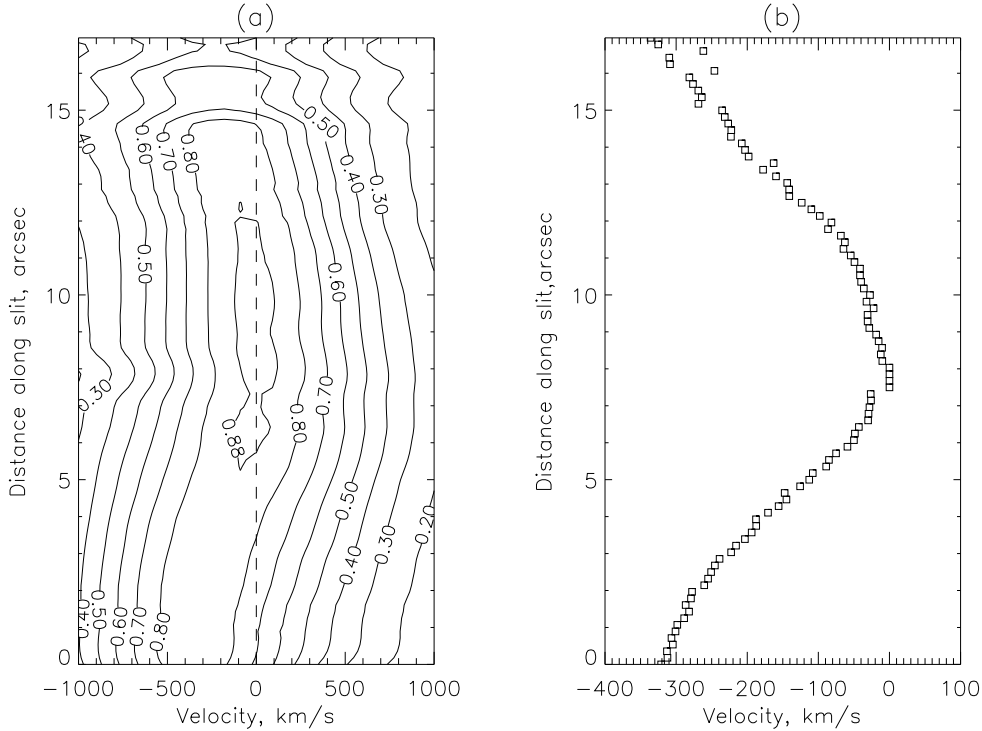}}
 \caption{Distortion of lines in the meteor spectrum taken
using  {\#}2. (a) Isophotes of the cross-correlation function computed for each
point along the slit height and (b) the radial-velocity curve along the slit
height.} \label{fig3:Afanasiev_n}
\end{figure*}

The distribution of geocentric velocities of meteors is known to have two peaks
located at about 30~km/s and in the 60--80~km/s interval [\cite{8:Afanasiev_n}].
The first peak is attributed to solar system particles, whereas the second peak
is attributed to particles captured by the Sun and moving in hyperbolic orbits.
Radar data are indicative of the excess of meteors in the domain of high
velocities (about 1{\%} of meteors have velocities above 100~km/s), however, no
meteor observations are known to yield velocities of several hundred km/s
[\cite{9:Afanasiev_n}]. Hence the question: what do we see?

The closest velocity --- the linear velocity of the rotation of the Sun about
the Galactic center --- is equal to about 220~km/s and is insufficient to
explain the meteor velocity we observed. An analysis of the motions of
Local-Group galaxies [\cite{10:Afanasiev_n}] shows that the Solar System moves at
a velocity of  316$\pm $11~km/s toward the mass center of the group in the
direction of $l=93\pm 2\degr$ and $b=-4\pm 1\degr$. Figure~4 compares the
position of the apex  of the motion toward the centroid of the Local Group with
the trajectory of the meteor on the sky. As is evident from the figure, the
agreement is remarkable: the radiant of the trajectory of our meteor coincides
with the position of the apex. Given the agreement of velocities, this fact
leads us to conclude that we observe an intergalactic particle, which is at
rest with respect to the mass centroid of the Local Group and which was ``hit''
by the Earth.

\begin{figure*}[htbp]
\centerline{\includegraphics[width=14cm]{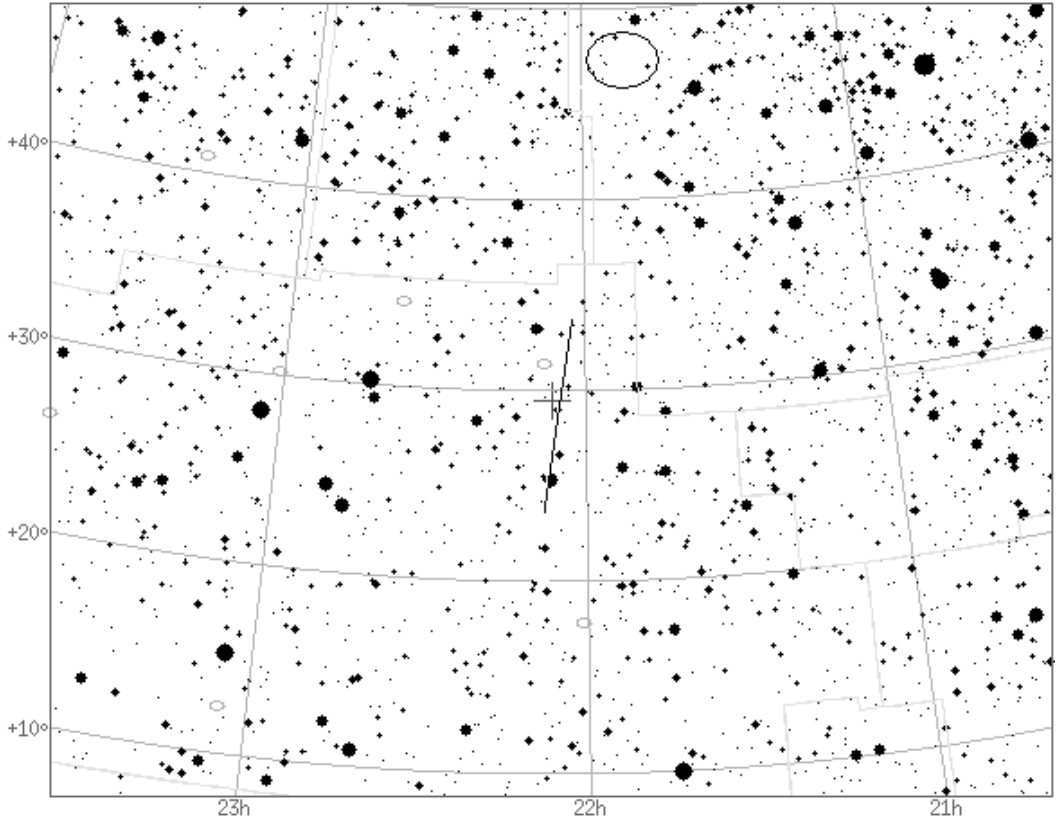}}
 \caption{ Portion of the sky map with the ellipse and
straight line showing the apex of motion toward the centroid of the Local Group
of galaxies and the track of the meteor recorded by the 6-m telescope of the
Special Astrophysical Observatory of the Russian Academy of Sciences,
respectively.} \label{fig4:Afanasiev_n}
\end{figure*}

\section{ESTIMATE OF THE SIZE AND MASS OF THE METEOR BODY}

When calibrated to the spectrophotometric standard, the integrated brightness
of the meteor estimated from the spectrum shown in Fig.~2 yields apparent
V-band magnitudes in the interval 9$\div12^m$ for the observed angular
velocities  1$\div $0.1\degr/s of the meteor. We have at our disposal
single-site observations of the event, making it impossible to obtain all
parameters of its trajectory that are required to correctly and completely
solve the inverse problem. We therefore may obtain only very coarse estimates.
The estimates of the apparent integrated magnitude of the meteor combined with
the velocity distribution along the line of sight imply that we are dealing
with the entry into the atmosphere of a rather small body moving at a velocity
of $v \ge 300$~km/s. Faint-meteor events usually occur at heights on the order
of 100~km and therefore we set $m$ equal to the absolute magnitude of the
meteor. We further assume that $v = 300$~km/s.

We now bring the equations of deceleration and mass loss of the
meteor body in the Earth's atmosphere to the following form:
$$
D\frac{dv}{dt} = - 0.5C_x \rho v^2
$$
\noindent
and
$$
\frac{dD}{dt} = - \frac{\Lambda }{6Q}\rho v^3,
$$
\noindent where
$$
D = M \mathord{\left/ {\vphantom {M S}} \right. \kern-\nulldelimiterspace}
S,
$$
\noindent $M$ and $S$ are the mass and area of the midsection (the maximum
section perpendicular to the direction of motion); $v$, its velocity; $\rho $,
the density of the atmosphere; $C_x $ and $\Lambda $, the resistance and heat
transfer coefficients, respectively; $Q$, the specific energy of destruction,
i.e., the energy required to carry unit mass away from the surface of the body
[\cite{11:Afanasiev_n}].

At such heights small bodies interact with the Earth's atmosphere in
free-molecule mode. Therefore, according to Kalenichenko~
[\cite{12:Afanasiev_n}], we must set  $\Lambda = 1$ and $C_x = 2$ in the above
equations. Here $Q \le 10^{11}$~erg/g, where the upper boundary corresponds, by
the order of magnitude, to evaporation of most of the compounds with the
allowance for heating  from absolute zero and for all the phase
transitions [\cite{12:Afanasiev_n}].

It is easy to verify that under such conditions the body must
disintegrate completely before its velocity appreciably decreases.
 We therefore estimate its size by integrating the
equation of mass loss at $v = const$ jointly with the equation of
the variation of the density of the atmosphere in the comoving
reference frame:
$$
\frac{d\rho }{dt} = \frac{\rho v\cos z}{H},
$$
\noindent where $z$ is the angle of entry of the body into the atmosphere
counted from the vertical direction (the zenith angle of the radiant of the
fireball); $H$, the isothermal atmosphere scale height in the Earth's gravity
field. Integrating from infinity  ($D = D_0 $, $\rho = 0)$ to the disintegration of the body
($D = 0$, $\rho = \rho )$, we obtain:

\begin{equation}
\label{eq1:Afanasiev_n}
D_0 = \frac{\Lambda H}{6Q\cos z}\rho v^2\,,
\end{equation}

\noindent i.e., a body of size $D_0 $ can penetrate into the
atmosphere down to the layers with density $\rho$. Values of atmosphere density $\rho$
have been calculated for each value of $h$ in accordance with the standard model atmosphere
[\cite{13a:Afanasiev_n}]. Formula (1)
can be used to estimate the lower limit of the characteristic
size of the body if we substitute into it the upper limit for
$Q$, because, according to Kalenichenko~[\cite{13:Afanasiev_n}],
$$
S = \kappa D^2.
$$

In our case we adopt $\kappa = 1$\,cm$^{6}$g$^{ -2}$, which is rather close to
the mode of this distribution for a substantially large sample of
nondisintegrating fireballs of the Peripheral network [\cite{14:Afanasiev_n}]
computed assuming that the scatter and shapes are the same for both the
fireballs and minor bodies. However, in this case it is evident that the value
$R_0 = \sqrt {\kappa D_0^2 } $  characterizes the size of the body. To make
estimates convenient, we also assume that $I$ (erg/s) obeys the well-known
(see, e.g.,~[\cite{15:Afanasiev_n}]) dependence on its absolute magnitude $m$
(the magnitude reduced to a distance of 100~km from the observer),

$$
\lg I = 9.72 - 0.4m\,,
$$

\noindent and is related to the loss of energy of the incident
gas via the midsection of the body by the following formula

$$
I = 0.5C_H S\rho v^3\,,
$$

\noindent where $C_H $ is the radiative heat transfer coefficient (see,
e.g., [\cite{13:Afanasiev_n}]), implying

$$
S = \frac{2I}{C_H \rho v^3}
$$
\noindent and characteristic size

$$
R = \sqrt S = \sqrt {\frac{2I}{C_H \rho v^3}}\,.
$$

No $C_H $ estimates are available for the Earth's atmosphere, not to mention
free-molecular flow, at such high hypersonic velocities. We take the liberty
to use the  $C_H \approx 10^{ - 3}$ value for Prairie Network fireball 40503
($v \approx 22$~km/s) at the point of its appearance at a height of about
95~km, where the flow about its body was still free molecular
[\cite{16:Afanasiev_n}].

The Table lists the estimates of the characteristic size of our meteor obtained
adopting all the above assumptions.

\begin{table}[htbp]

{\bf Estimates of the characteristic size of the meteor body}
\label{tab1:Afanasiev_n}
\medskip

\begin{tabular}{l|l|l|l}
\hline $h,\mbox{k}\mbox{m}$& $R_0 > ,\mbox{c}\mbox{m}$& $R,\mbox{c}\mbox{m}(m =
9)$&
$R,\mbox{c}\mbox{m}(m = 12)$ \\
\hline
100&
1&
0.01&
0.004 \\
\hline
110&
0.2&
0.03&
0.01 \\
\hline
120&
0.05&
0.06&
0.02 \\
\hline
\end{tabular}
\end{table}

\begin{figure*}[tbp]
\centerline{\includegraphics[width=14cm]{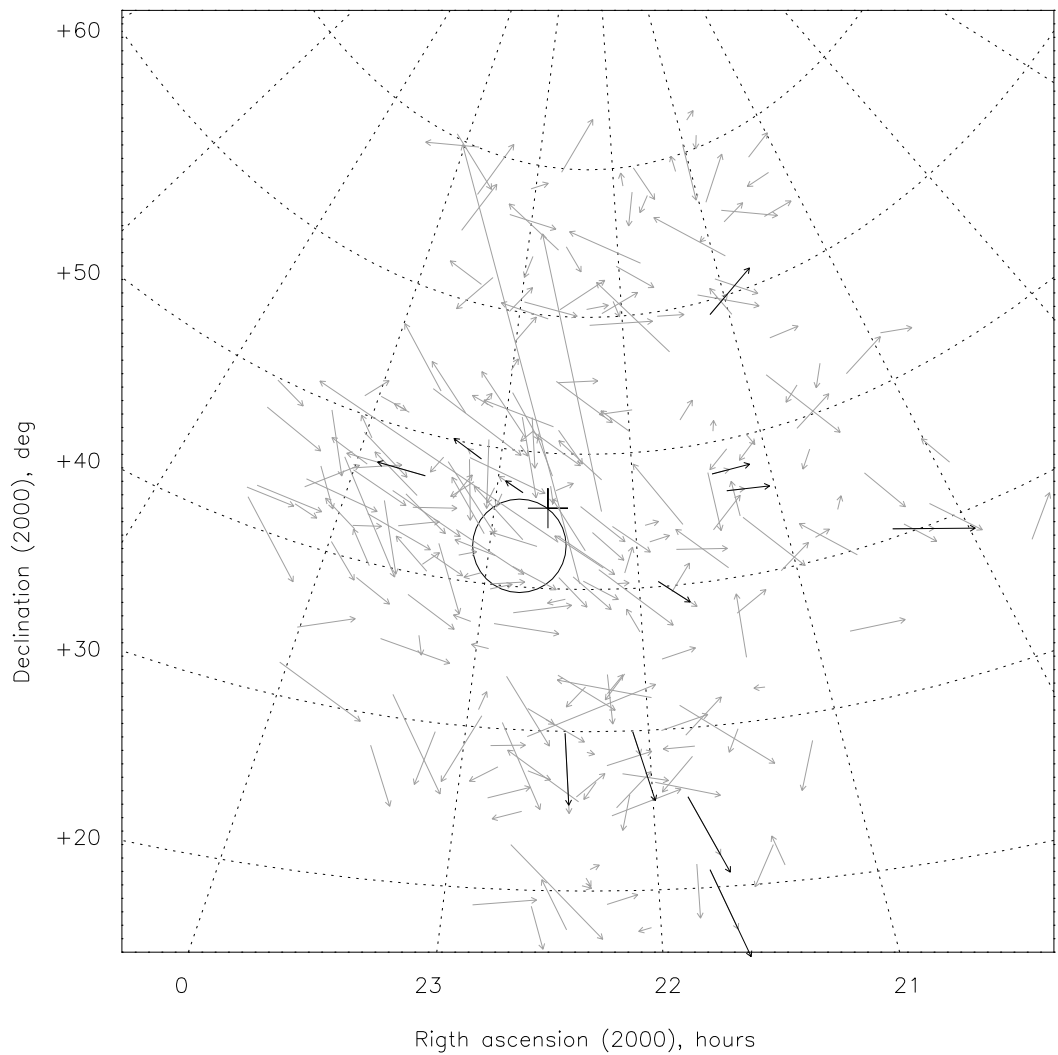}}
 \caption{Tracks of meteors recorded with FAVOR wide-angle
camera. The black arrows indicate the meteors emerging from the radiant, which
we marked by the cross, and the circle shows the position of the apex of
motion toward the centroid of the Local Group of galaxies.}
\label{fig5:Afanasiev_n}
\end{figure*}

\begin{figure*}[htbp]
\centerline{\includegraphics[width=14cm]{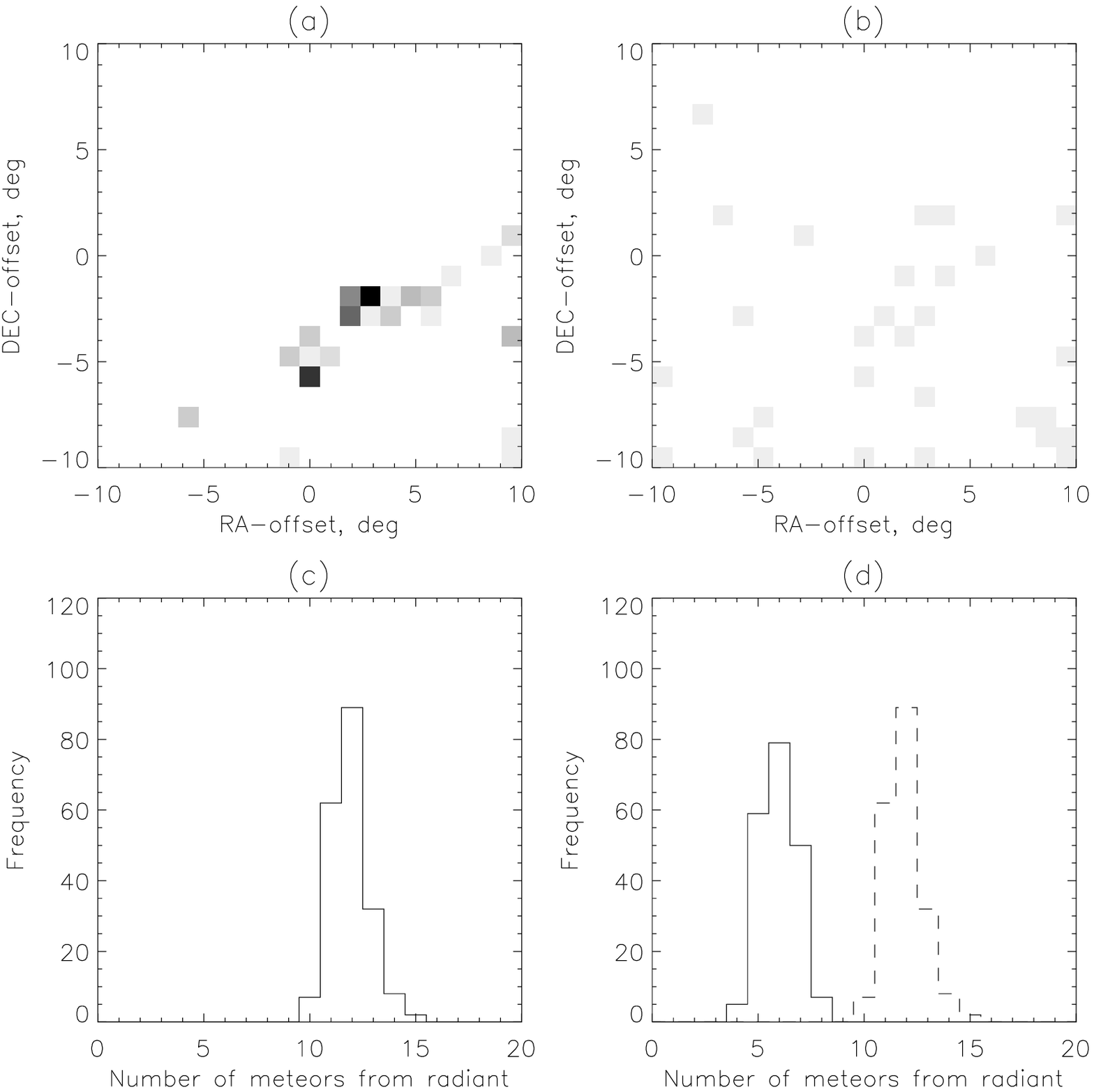}}
 \caption{Distribution of the occurrence frequency of
meteor tracks satisfying the condition of emergence from the radiant with the
given coordinates. (a) The observed two-dimensional distribution of the
occurrence frequency of meteors for different coordinates of the radiant with
respect to the center with RA 22$^{h}$20$^{m}$ and DEC +48$^{^{\circ}}$, (c)
the distribution of the observed number of meteors satisfying the condition of
emergence from the radiant. (b) and (d) Same as figures (a) and (c), but for
the model of randomly distributed track orientations. The dashed line shows the
observed distribution of the number of tracks to be compared with the model
shown in the figure (d).} \label{fig6:Afanasiev_n}
\end{figure*}

The size of the body can also be estimated in a different way. The meteor moved
by at least 3 arcmin during the time while it remained in the field of view of
the camera (Fig.\,1). This displacement $l$ corresponds to about 100~m, or two
orders of magnitude less than $H$, at a height of 100~km. In this case we can
neglect the change in $\rho $ over distance  $l$ and use the following relation
instead of  (\ref{eq1:Afanasiev_n}):

\begin{equation}
\label{eq2:Afanasiev_n} D_0 = \frac{\Lambda }{6Q}\rho v^2l\,.
\end{equation}

\noindent This relation yields, under all the above assumptions, the size $R_0$
of the meteor body on the order of 0.01~cm at a height of 100~km (compare with
the values listed in the Table). The result also confirms that our adopted
$C_H$ value is sufficiently correct. Note that substituting $Q = 10^{11}$~erg/g
into formulas (\ref{eq1:Afanasiev_n}) and (\ref{eq2:Afanasiev_n}) yields
underestimated  $R_0 $ values, and the same is true for the result obtained by
substituting the lower boundary for $l$ into formula (\ref{eq2:Afanasiev_n}).
The actual size of the body may be several times greater than that.

One may expect that the time scale of the disintegration of a particle moving
in the Earth's atmosphere at a velocity on the order of 300~km/s is less than
0.1~s, and therefore the observed meteor may have an integrated magnitude of
more than 9 magnitude. If particles disintegrate at higher altitudes, then their
sizes must be greater all other conditions being the same. Therefore the
characteristic size listed in the table is a lower estimate. We adopt a burnout
height of $h>100$~km to infer a size estimate of $R>10^{-2}$~cm. If the observed
meteor particle is a chondrite (with a density of about 3 g/cm$^{3})$, then the
mass of the particle can be estimated at M$>7\cdot 10^{-6}$~g.

Note that our estimate, which we obtained under rather coarse assumptions,
agrees with numerical computations of the expected parameters (magnitudes and
masses) of interstellar meteors moving with velocities as high as 500~km/s
[\cite{17:Afanasiev_n}]. Hill et al. [\cite{17:Afanasiev_n}] showed that meteors
with the velocities of 300~km/s and masses on the order of 10$^{-5}$~g must be
observed as about 9$^m$ objects disintegrating at the heights of 110--120~km.

It is interesting that our estimate coincides with the size of chondrules in
carbonaceous chondrites (see, e.g., [\cite{18:Afanasiev_n})], which are usually
smaller than 0.5~mm and constrained to a rather narrow interval. Many
chondrules have spherical or spheroidal shapes and appear to have formed as a
result of fast crystallization and hardening of melt drops. The distribution of
carbon isotopes in carbonaceous chondrites differs widely from that of
terrestrial and lunar rocks and is suggestive of extrasolar and (or) presolar
(pregalactic) material.

\section{ESTIMATE OF THE NUMBER OF HIGH-VELOCITY METEORS}

By itself, the fact of observing a high-velocity meteor provides no information
about the number of such particles, and the coincidence of the direction of its
trajectory with the direction toward the apex of motion toward the centroid of
the Local Group of galaxies does not prove the extragalactic nature of the
meteor. To estimate the number of high-velocity meteors and search for the
position of their radiant near the apex, we observed in October--November, 2006
faint sporadic meteors with FAVOR wide-angle CCD camera developed at the
Special Astrophysical Observatory of the Russian Academy of Sciences and
Federal Research Institute of Semiconductor Devices  [\cite{19:Afanasiev_n}]. The
camera has a 16$\times22\degr$ field of view; pixel size of 1.5~acrmin;
single-frame exposure of 0.12~s; limiting V-band magnitude of 10.5, and a
signal-to-noise ratio of  5.

We conducted observations in six fields located around the apex for a total
time of 34.5 hours during six observing nights to take about 10$^{6}$ frames
from which we selected about 2000 ones with recorded meteors. After astrometric
calibration we determined the coordinates of the beginning and end of the
meteor tracks. The track morphology and repeated track images  (for meteors
moving with low angular velocities) allowed us to determine the direction of
the meteor motion on the sky. We identified a total of 246 meteors with a mean
magnitude of V$\approx7^{m}$ in the field studied. The magnitudes of the
faintest of these meteors lied in the 9--9.5$^{m}$ interval. The median angular
velocity of the meteors found is equal to 0.8\degr/s. We show their tracks as
gray dashes in Fig.\,5.

To verify whether the field of recorded meteors had a radiant, we tested the
coincidence of the direction of the radius vector pointing from the given
radiant point to the start of the track with the direction of the apparent
trajectory of the meteor. We varied the position of the radiant within $\pm
$10\degr in both coordinates about the apex position. The size of the error box
for the track direction is about 3 degrees and it is determined by the accuracy
of astrometric calibration (about 20$\arcsec$). The left-hand panel in Fig.\,6
shows the distribution of the number of meteors that obey the condition of
emerging from the radiant and the observed density of the occurrence frequency
of such meteors in the sky plane. It is evident from the figure that about 12
meteors out of 246 obey the condition of motion from the radiant, and the
position of the radiant can be confidently determined from the maximum of the
distribution function of the occurrence frequency in the sky plane. For a
$3\degr$  error box the expected number of meteors should be equal to about three
in the case of a random distribution of coordinates and orientations of tracks.
In our field the distribution of track coordinates is not uniform and to test the
statistical significance of the result obtained, we numerically simulated tracks
with the coordinates coinciding with those of the observed tracks and
accidentally distributed directions (with uniformly distributed angles).
Figure~6 shows the result of our estimate of the number of meteors satisfying
the condition of motion from the radiant for the case of accidental
distribution of track directions. It follows from this figure that the number
of such meteors is equal to about five for a random model, but there is no
distinguished point in the distribution of the occurrence frequency on the sky
plane. Judging by the form of the distributions of observed track directions
for the random-distribution model for the observed track directions, our radiant
located near the apex of the motion toward the centroid of the Local Group of
galaxies is not accidental at a confidence level of >96{\%}. In Fig.\,5 the
black arrows, circle, and cross show the meteor tracks satisfying the condition
of motion from the radiant, the position of the apex corrected for the motion
of the Earth, and the inferred position of the radiant, respectively.

Given the distance between the track and the radiant and the track length, one
can estimate the geocentric velocity of the meteors. We obtain a velocity
estimate of $282\pm$53~km/s for a 110~km intrusion height into the Earth's
atmosphere and an average magnitude of 8.2. Note that the estimates of the
velocity and brightness of meteors so obtained agree with spectroscopic data.

\section{ESTIMATE OF THE SPACE DENSITY OF INTERGALACTIC DUST}

Observations of the flares of high-velocity particles in the
upper layers of the atmosphere allow us to estimate the space
density $n_{d}$ of such particles. Given that the velocity
$V_{d}$ of fast dust particles exceeds the velocity of the Earth
by one order of magnitude, space density $n_{d}$ can be estimated
as

$$n_{d} =(S_{0}V_{d}t)^{-1}\,,$$

\noindent where $S_{0}$ is the survey area during the search for events
(flares), and $t $ is the average time interval between observed events. In our
case for the patrol field of view with the size of about $40\degr$ the survey
area at the altitude of $\sim $110~km and  a mean zenith distance of
$z\approx 40\degr$ is equal to $S_{0}\approx 10^{14}$~cm$^{2}$. The data
reported above allow us to adopt t$\approx 3^{h}\approx 10^{4}$~s as the
average time interval between events in this field. For a velocity of
$V_{d}\approx $300~km/s we obtain \mbox{$n_{d}=3.6\cdot10^{-26}$~cm$^{-3}$.}
For a typical dust-grain mass of \mbox{$M_{d}\ge 0.7\cdot10^{-5}$~g,} we find
the average density of intergalactic particles in the Earth's vicinity
\mbox{$\rho \ge 2.5\cdot 10^{-31}$~g/cm$^{3}$,} i.e., about  2.5{\%} of the
critical density.

Let us assume that extragalactic dust with the above density is uniformly
distributed over the entire volume of the Local Group. According to
Karachentsev~[\cite{20:Afanasiev_n}], Local Group of galaxies can be
characterized by its ``zero velocity sphere'' radius $R_{LG} =$ \mbox{$(1\pm 0.1)$~~Mpc}~~~ and
~~~total~~~ mass~~~ \mbox{$M_{LG}=$} \mbox{$(1.3\pm 0.3)\cdot 10^{12}M_{\sun}$.} The total mass of
dust in this volume is equal to \mbox{$1.6\cdot10^{10}M_{\sun}$}, or about 1{\%}
of the total mass of the group. It does not appear unusual, because such a mass
has no effect on the dynamics of the Local Group of galaxies.

On the other hand, the intergalactic dust medium may show up in
integrated or selective absorption. Let us assume that this
medium with density $n_{d}$ and characteristic dust-grain radius
of $R_{d}\sim 10^{-2}$~cm uniformly fills the Universe out to the
horizon R$_{c}$=c/H$_{0}$, where c is the speed of light and
H$_{0}$, the Hubble constant (72~km/s/Mpc). In this case the
optical depth of the dusty Universe is
$$
\tau = \pi R_d^2 n_d c / H_0\,,
$$
\noindent if we neglect various evolutionary effects. The above
estimate of $n_{d}$ = 3.6$ \cdot $10$^{-26}$~cm$^{-3}$ implies an
optical depth of the Universe of only  \textit{$\tau $ }$ \approx
$ 0.15, which would have no apparent effect on the spectra of
distant galaxies and quasars. In reality, one may expect that
intergalactic dust should concentrate in groups and clusters of
galaxies. According to Karachentsev and
Lipovetsky~[\cite{21:Afanasiev_n}], the optical depth of clusters
of galaxies does not exceed $\tau \approx $0.2, implying no
conflicts with observational data.

Of course, the following important questions remain open:

(1) How have metal-rich dust particles come to the intergalactic
space? In principle, various mechanisms may explain this ---
supernova explosions, tidal interactions of galaxies, radiation
pressure, condensation in molecular clouds (chondrules?), etc.

(2) Why are the sizes of intergalactic particles  two orders
smaller (and their masses are six orders of magnitude greater)
than common interstellar dust particles in our Galaxy?

(3) If intergalactic dust associates with galaxies in the form of dust
aureoles, is it possible now to find such formations (IR observations onboard
Spitzer)? As is well known, the aureoles of ionized gas around nearby spiral
galaxies are already observed [\cite{22:Afanasiev_n}].

(4) If nonuniform intergalactic dust medium with the above
parameters does exist, would it show up as irregularities on
cosmic microwave radiation (WMAP etc.)?

We expect to answer these questions in the nearest future.

\begin{acknowledgements} 
We are grateful to E.~Katkova for performing observations with
FAVOR camera, S.~Karpov for  preliminary reduction of
real-time data, and to R.Z.~Sag\-deev and A.M.~Fridman for
fruitful discussions.
\end{acknowledgements}

\end{document}